\documentclass[aip,graphicx,jcp]{revtex4}

\begin{document}


\title{Contact theorems for anisotropic fluids near a hard wall.} 



\author{M. Holovko}
\affiliation{Institute for Condensed Matter Physics,
National Academy of Sciences, 1 Svientsitskii Str., 79011 Lviv, Ukraine\\}
\author{D. di Caprio}
\affiliation{Institute of Research of Chimie Paris, CNRS - Chimie ParisTech,\\
11, rue P. et M. Curie, 75005 Paris, France\\}



\begin{abstract}
In this paper, starting from the Born-Green-Yvon (BGY) equation,
we derive a general expression for the contact value of the singlet
distribution function near a hard wall for anisotropic fluids.
This relation includes two separate contributions. One is connected to the
partial bulk pressure relative to a given orientation of the molecules.
The second one is connected to the anchoring phenomena and is characterized
by the direct interaction between the molecules and the wall.
From this relation, we then formulate the contact theorem for the density and
the order parameter profiles. The results are illustrated on the case of a
nematic fluid near a hard wall.
\end{abstract}


\maketitle 

\section{Introduction}
The contact theorem for a fluid near a hard wall is one of the few exact theoretical
relations describing a fluid at an interface. According to this theorem, for a simple
fluid, the contact value of the density profile $\rho(z)$ near a hard wall at position
$z=0$ (the wall) is determined by the bulk pressure $P$ of the fluid
\begin{eqnarray} \label{eq:CT0}
  \rho(z=0) = \beta P
\end{eqnarray}
where $z$ is the normal distance from the wall and $\beta = 1/ (k_B T)$
with $k_B$ the Boltzmann constant and $T$ the absolute temperature.

In the case of ionic fluids near a charged hard wall, eq. (\ref{eq:CT0}) should
be corrected by adding to the pressure $P$ the electrostatic Maxwell tensor
contribution \cite{HendersonBlumJCP1978,HendersonBlumLebowitz1979}.
The contact theorem for the charge profile for ionic fluids near a charged hard
wall was also formulated
\cite{HolovkoBadialidiCaprio2005,HolovkoBadialidiCaprio2007,HolovkodiCaprio2008}.
It was shown that in contrast to the contact theorem for the density profile,
the contact theorem for the charge profile has a non local character.
In particular for the symmetrical electrolyte, the contact value of the charge
profile is expressed by the integral of the product of the density profile and the
gradient of the mean electric potential.

In contrast to simple and ionic fluids, in anisotropic fluids such as nematic
liquid crystals or anisotropic dipolar fluids, the singlet distribution function
depends not only from the distance $z$ between the wall and the particles but
also on the orientations of the molecules (for instance molecule 1) with respect to the preferred
orientational direction (director $n$) $\Omega_{1n} = (\theta_{1n},\phi_{1n})$
and from the orientation of the wall with respect to this director $\Omega_{wn}$.
Due to the orientational ordering, anisotropic fluids near a hard wall have a far
richer behaviour than simple and ionic fluids.
Between them we can mention the tensorial character of the order parameter,
the wall-induced biaxyality and the anchoring phenomena, whereby the surface
induces a specific orientation of the director with respect to the surface
orientation~\cite{Jerome1991}.
In this study, we consider anisotropic fluids with only orientational ordering
excluding for instance smectic type phases where spatial ordering is also
present.

In this paper, in a similar way as for ionic fluids
\cite{HolovkoBadialidiCaprio2005,CarnieChan1981}, starting from
the Born-Green-Yvon equation for the singlet distribution
function $\rho(z,\Omega_{1n},\Omega_{wn})$, we formulate the contact theorem
for an anisotropic fluid near a hard wall.
From the obtained relation, we will the formulate the contact theorems
for the density profile
\begin{eqnarray} \label{eq:rhointegW}
  \rho(z,\Omega_{wn}) \equiv \int d\Omega_{1n}\; \rho(z,\Omega_{1n},\Omega_{wn})
\end{eqnarray}
and for the order parameter profile
\begin{eqnarray} \label{eq:SintegW}
  S_{l\mu} (z,\Omega_{wn}) \equiv \int d\Omega_{1n}\; \rho(z,\Omega_{1n},\Omega_{wn})\, Y_{l\mu}(\Omega_{1n})
\end{eqnarray}
where $d\Omega_{1n}=\frac{1}{4\pi} sin(\theta_{1n})d\theta_{1n}d\phi_{1n}$
is the normalized angle element of the orientation of the molecules
with respect to the director, $Y_{l\mu}(\Omega_{1n})$ is the corresponding
spherical harmonic for the orientation of the molecule with respect to the director
\footnote{Note that $S_{l\mu} (z,\Omega_{wn})$ for $\mu \neq 0$ is complex,
it is usual to introduce the following real quantities $S_{l\mu}^+ = (S_{l\mu} + S_{l\mu}^*)/2$
and $S_{l\mu}^- = (S_{l\mu} - S_{l\mu}^*)/(2i)$ where * denotes the complex conjugate.}.
Usually the important spherical harmonics are such that $l=1$ for dipolar fluids
and $l=2$ for nematic fluids.

\section{Derivation of the contact theorem}
In order to take into account anchoring effects, we assume that the molecules
can interact with the wall with the potential $V(z,\theta_{1w})$ where the
dependence is on the distance $z$ to the wall and on
$\theta_{1w}$ the angle between the molecule and the wall.
As a result the BGY equation for anitropic fluids near a hard wall is written
\begin{eqnarray} \label{eq:BGY0}
  \nabla_1 \rho(1) = - \beta \nabla_1 V(1) \rho(1) - \beta
    \rho(1) \int d2\; \rho(2) g(12) \nabla_2 U(12)
\end{eqnarray}
where $1$ and $2$ stand for the position and orientation of the corresponding
molecules and $U(12) = U(r_{12}, \Omega_{1n}, \Omega_{2n})$
is the potential of the intermolecular interaction, $r_{12}$
denotes the distance between the molecules $1$ and $2$ and $\Omega_{1n}, \Omega_{2n}$
are the respective orientations of the molecules and $g(12)$
is the pair distribution function.

After integration of eq. (\ref{eq:BGY0}) with respect to $z$, the inhomogeneous
part of the pair distribution function $g^{inh}(12)$ vanishes as seen
in \cite{HolovkoBadialidiCaprio2005,CarnieChan1981} due to the symmetry of the pair
interaction potential which corresponds to $\nabla_1 U(12) =- \nabla_2 U(12)$.
As a result, we have
\begin{eqnarray} \label{eq:CT1}
  \rho(z=0,\Omega_{1n},\Omega_{wn}) &=& \beta \int_{0}^{\infty} dz_1 \;
    \frac{\partial V(z_1,\Omega_{1n},\Omega_{wn})}{\partial z_1}
    \rho(z_1,\Omega_{1n},\Omega_{wn})\nonumber\\
  &+& \rho_b(\Omega_{1n}) \left(1 -\frac{1}{6}\int d\mathbf{r}_{12}
  d\Omega_{2n} \;g_b(r_{12},\Omega_{1n},\Omega_{2n}) \rho_b(\Omega_{2n})r_{12}
   \frac{\partial U(12)}{\partial r_{12}}  \right)
\end{eqnarray}
where $\rho_b(\Omega_{1n})$ and $g_b(r_{12},\Omega_{1n},\Omega_{2n})$
are correspondingly the singlet and pair distribution functions for the
anisotropic fluid in the bulk phase.

The last term in eq. (\ref{eq:CT1}) can be treated as the bulk partial pressure
for molecules with a given orientation $\Omega_{1n}$
\begin{eqnarray}
  \beta P(\Omega_{1n}) \equiv \rho_b(\Omega_{1n}) \left(1
-\frac{1}{6}\int d\mathbf{r}_{12}d\Omega_{2n}\; g_b(r_{12},\Omega_{1n},\Omega_{2n}) \rho_b(\Omega_{2n})r_{12}
   \frac{\partial U(12)}{\partial r_{12}}\right)
\end{eqnarray}
As a result the contact theorem for the singlet distribution function
$\rho(z,\Omega_{1n},\Omega_{wn})$ can be presented in the form
\begin{eqnarray} \label{eq:CT2}
  \rho(z=0,\Omega_{1n},\Omega_{wn}) &=& \beta \int_{0}^{\infty} dz_1\;
    \frac{\partial V(z_1,\Omega_{1n},\Omega_{wn})}{\partial z_1}
    \rho(z_1,\Omega_{1n},\Omega_{wn}) + \beta P(\Omega_{1n})
\end{eqnarray}
This constitutes the main result of this paper.

As we can see the dependence of the contact value of the singlet distribution
function from the angle between the wall and the director: $\Omega_{wn}$ appears
only in the first term accounting for the direct interaction between the wall
and the particles.
In the absence of this interaction the contact value of the singlet
distribution function does not depend on $\Omega_{wn}$ and
is set only by the partial pressure, hence
\begin{eqnarray}\label{eq:CT2b}
  \rho(z=0,\Omega_{1n},\Omega_{wn}) = \beta P(\Omega_{1n})
\end{eqnarray}
This means that we can omit $\Omega_{wn}$ in eq. (\ref{eq:CT2b}).

After integration of eq. (\ref{eq:CT2}) on all possible orientations of molecule $1$, that
is integration over $\Omega_{1n}$, using the definition eq. (\ref{eq:rhointegW}),
we obtain the contact theorem for the density profile
\begin{eqnarray} \label{eq:CT3}
  \rho(z=0,\Omega_{wn}) = \beta \int_{0}^{\infty} dz_1 d\Omega_{1n}\;
    \frac{\partial V(z_1,\Omega_{1n},\Omega_{wn})}{\partial z_1}
    \rho(z_1,\Omega_{1n},\Omega_{wn}) + \beta P
\end{eqnarray}
where
\begin{eqnarray}
   \beta P = \int d\Omega_{1n}\; P(\Omega_{1n})
\end{eqnarray}
is simply the bulk pressure for the anisotropic fluid.

We shall mention that in the absence of anchoring effects i.e. when
$V(z,\Omega_{1n},\Omega_{wn})=0$ the contact theorem
eq. (\ref{eq:CT3}) coincides exactly
with the contact theorem eq. (\ref{eq:CT0}) for the
density profile in the case of a simple fluid.
As we have already mentionned, in this case the contact value
$\rho(z=0,\Omega_{wn})$ is independent from the angle $\Omega_{wn}$ between the
wall and the director.
This result was previously illustrated in \cite{HolovkoKravtsivdiCaprio2013} in the
framework of the mean field approximation (MFA) for a nematic fluid near a hard
wall in the Maier-Saupe model \cite{MaierSaupe1959,MaierSaupe1960}.

Multiplying eq. (\ref{eq:CT2}) by $Y_{l\mu}(\Omega_{1n})$,
integrating over $\Omega_{1n}$ and using definition
eq. (\ref{eq:SintegW}), we can formulate the contact theorem
for the order parameter tensor
\begin{eqnarray} \label{eq:CTS0}
  S_{l\mu} (z=0,\Omega_{wn}) &=& \beta \int_{0}^{\infty} dz_1 d\Omega_{1n}\;
    Y_{l\mu}(\Omega_{1n}) \frac{\partial V(z_1,\Omega_{1n},\Omega_{wn})}{\partial z_1}
    \rho(z_1,\Omega_{1n},\Omega_{wn})\nonumber\\
   &&+ \beta \int d\Omega_{1n} \;Y_{l\mu}(\Omega_{1n}) P(\Omega_{1n})
\end{eqnarray}

In the absence of anchoring interaction
$V(z,\Omega_{1n},\Omega_{wn})=0$, we have the simpler relation
\begin{eqnarray} \label{eq:CTS1}
  S_{l\mu} (z=0,\Omega_{wn}) = \beta \int d\Omega_{1n}\; Y_{l\mu}(\Omega_{1n}) P(\Omega_{1n}).
\end{eqnarray}
For this case $S_{l\mu} (z=0,\Omega_{wn})$ does again not depend from the angle
$\Omega_{wn}$ and if $P(\Omega_{1n})$ has no azimuthal dependence
\begin{eqnarray} \label{eq:CTS2}
  S_{l\mu} (z=0,\Omega_{wn}) = 0  \hspace{2ex}\mathrm{for}\hspace{1ex}\mu \neq 0.
\end{eqnarray}

\section{The case of nematic fluids}
To illustrate the role of anchoring on the contact theorems, we consider a
simple model of nematic fluid near a hard wall with the wall-nematic interaction
introduced in \cite{Sokolovska2004,Sokolovska2005}
\begin{eqnarray}
  V(z,\Omega_{1n},\Omega_{wn}) = V_0(z) + V_2(z) P_2(cos[\theta_{1w}])
\end{eqnarray}
where the first term and second term describe respectively the direct isotropic and anisotropic
interactions of the molecules with the wall
and $P_2(cos(\theta_{1w})) = (3cos^2[\theta_{1w}]-1)$
is the second order Legendre polynomial.

As a result, following relations (\ref{eq:CT3}) and (\ref{eq:CTS0})
and the definitions (\ref{eq:rhointegW}) and (\ref{eq:SintegW}),
the contact theorems for the density profile and the order parameter
can be presented as
\begin{eqnarray}
  \rho(z=0,\Omega_{wn}) =&& \beta \int_{0}^{\infty} dz_1
    \frac{\partial V_0(z_1)}{\partial z_1} \rho(z_1,\Omega_{wn})\nonumber\\
   &+&\beta \sum_\mu Y_{2\mu}(\Omega_{wn}) \int_{0}^{\infty} dz_1 \frac{\partial V_2(z_1)}{\partial z_1} S_{2\mu} (z_1,\Omega_{wn})\nonumber\\ &+&\beta P\\
  S_{2\mu}(z=0,\Omega_{wn}) =&& \beta \int_{0}^{\infty} dz_1
    \frac{\partial V_0(z_1)}{\partial z_1} S_{2\mu}(z_1,\Omega_{wn}) \nonumber\\
   &+&\beta \sum_{\mu'} Y_{2\mu'}(\Omega_{wn}) \int_{0}^{\infty} dz_1 \frac{\partial V_2(z_1)}{\partial z_1} S_{2\mu\mu'} (z_1,\Omega_{wn})\nonumber\\ &+&\beta \int d\Omega_{1n} Y_{2\mu}(\Omega_{1n}) P(\Omega_{1n})
\end{eqnarray}
where
\begin{eqnarray}
  S_{2\mu\mu'} (z,\Omega_{wn}) = \int d\Omega_{1n}\; \rho(z,\Omega_{1n},\Omega_{wn}) Y_{2\mu}(\Omega_{1n})
      Y_{2\mu'}(\Omega_{1n})
\end{eqnarray}
As we can see in contrast to the contact theorem eq. (\ref{eq:CT2}) for the singlet distribution
function the presence of a orientationnally dependent interaction between molecules
and wall responsible for the anchoring phenomenon, leads to a non closed
set of integral equations for the contact theorem for the density and the order
parameters profiles.
That is for the calculation of the contact value of the density profile, the knowledge of the density and
order parameter profiles are necessary.
And for the calculation of the contact value of the order parameter, the
knowledge of the order parameter profile as well as the higher order parameter
$S_{2\mu\mu'} (z,\Omega_{wn})$ profile are needed.

\section{Conclusion.}
In this paper, we give the exact expression for the contact value
of the singlet distribution function near a hard wall for an anisotropic fluid
which can be characterized by an orientational ordering in the bulk.
The expression includes two separate contributions. One is connected with the partial
pressure $P(\Omega_{1n})$ relative to a given orientation $\Omega_{1n}$ of the molecules.
The second contribution is connected with the direct interaction
of the molecules with the wall which can be distance and orientationnally
dependent which is the typical case of the so called anchoring phenomena.\\
Using this relation, we then formulate the contact theorem for the density
and the order parameter irrespective of the molecule orientation.
We show the fundamental role of the orientational dependent interaction
of the molecules with the wall on the tensorial character of the
order parameter near the wall.
In the absence of molecule-wall interaction, the contact values of the density
and of the order parameter profiles do not depend from the angle
between the wall and the director.
Then for dipolar or nematic liquids for instance, since the partial pressure
$P(\Omega_{1n})$ has no azymuthal orientational dependence, the contact values
of the singlet distribution function also will not have an azymuthal dependence.
As a result, the contact value of the order parameter will not have
an azymuthal component for the spherical harmonics with $\mu\neq0$.
Finally, we illustrate the contact theorem on a simple model of a nematic
fluid near a hard wall.
The set of equations has a non closed form.\\
We believe that these exact relations at a surface can be used to verify the
accuracy of results from numerical simulations or else improve density profile
approximations using state of the art approximations for the bulk pressure which
is easier to calculate.
We have used such a procedure for isotropic fluids
\cite{KravtsivPatsahanHolovkodiCaprio2015} and intend to apply to nematic fluids.

\begin{acknowledgments}
This work has been carried out in the framework of the PICS agreement
between the french Centre National de la Recherche
Scientifique (CNRS) and the National Academy of Sciences of Ukraine (NASU).
\end{acknowledgments}

\bibliography{paper}

\begin{thebibliography}{14}%
\makeatletter
\providecommand \@ifxundefined [1]{%
 \@ifx{#1\undefined}
}%
\providecommand \@ifnum [1]{%
 \ifnum #1\expandafter \@firstoftwo
 \else \expandafter \@secondoftwo
 \fi
}%
\providecommand \@ifx [1]{%
 \ifx #1\expandafter \@firstoftwo
 \else \expandafter \@secondoftwo
 \fi
}%
\providecommand \natexlab [1]{#1}%
\providecommand \enquote  [1]{``#1''}%
\providecommand \bibnamefont  [1]{#1}%
\providecommand \bibfnamefont [1]{#1}%
\providecommand \citenamefont [1]{#1}%
\providecommand \href@noop [0]{\@secondoftwo}%
\providecommand \href [0]{\begingroup \@sanitize@url \@href}%
\providecommand \@href[1]{\@@startlink{#1}\@@href}%
\providecommand \@@href[1]{\endgroup#1\@@endlink}%
\providecommand \@sanitize@url [0]{\catcode `\\12\catcode `\$12\catcode
  `\&12\catcode `\#12\catcode `\^12\catcode `\_12\catcode `\%12\relax}%
\providecommand \@@startlink[1]{}%
\providecommand \@@endlink[0]{}%
\providecommand \url  [0]{\begingroup\@sanitize@url \@url }%
\providecommand \@url [1]{\endgroup\@href {#1}{\urlprefix }}%
\providecommand \urlprefix  [0]{URL }%
\providecommand \Eprint [0]{\href }%
\providecommand \doibase [0]{http://dx.doi.org/}%
\providecommand \selectlanguage [0]{\@gobble}%
\providecommand \bibinfo  [0]{\@secondoftwo}%
\providecommand \bibfield  [0]{\@secondoftwo}%
\providecommand \translation [1]{[#1]}%
\providecommand \BibitemOpen [0]{}%
\providecommand \bibitemStop [0]{}%
\providecommand \bibitemNoStop [0]{.\EOS\space}%
\providecommand \EOS [0]{\spacefactor3000\relax}%
\providecommand \BibitemShut  [1]{\csname bibitem#1\endcsname}%
\let\auto@bib@innerbib\@empty
\bibitem [{\citenamefont {Henderson}\ and\ \citenamefont
  {Blum}(1978)}]{HendersonBlumJCP1978}%
  \BibitemOpen
  \bibfield  {author} {\bibinfo {author} {\bibfnamefont {D.}~\bibnamefont
  {Henderson}}\ and\ \bibinfo {author} {\bibfnamefont {L.}~\bibnamefont
  {Blum}},\ }\href@noop {} {\bibfield  {journal} {\bibinfo  {journal} {J. Chem.
  Phys.}\ }\textbf {\bibinfo {volume} {69}},\ \bibinfo {pages} {5441} (\bibinfo
  {year} {1978})}\BibitemShut {NoStop}%
\bibitem [{\citenamefont {Henderson}, \citenamefont {Blum},\ and\ \citenamefont
  {Lebowitz}(1979)}]{HendersonBlumLebowitz1979}%
  \BibitemOpen
  \bibfield  {author} {\bibinfo {author} {\bibfnamefont {D.}~\bibnamefont
  {Henderson}}, \bibinfo {author} {\bibfnamefont {L.}~\bibnamefont {Blum}}, \
  and\ \bibinfo {author} {\bibfnamefont {J.}~\bibnamefont {Lebowitz}},\
  }\href@noop {} {\bibfield  {journal} {\bibinfo  {journal} {J. Electroanal.
  Phys.}\ }\textbf {\bibinfo {volume} {102}},\ \bibinfo {pages} {315} (\bibinfo
  {year} {1979})}\BibitemShut {NoStop}%
\bibitem [{\citenamefont {Holovko}, \citenamefont {Badiali},\ and\
  \citenamefont {di~Caprio}(2005)}]{HolovkoBadialidiCaprio2005}%
  \BibitemOpen
  \bibfield  {author} {\bibinfo {author} {\bibfnamefont {M.}~\bibnamefont
  {Holovko}}, \bibinfo {author} {\bibfnamefont {J.}~\bibnamefont {Badiali}}, \
  and\ \bibinfo {author} {\bibfnamefont {D.}~\bibnamefont {di~Caprio}},\
  }\href@noop {} {\bibfield  {journal} {\bibinfo  {journal} {J. Chem. Phys.}\
  }\textbf {\bibinfo {volume} {123}},\ \bibinfo {pages} {234705} (\bibinfo
  {year} {2005})}\BibitemShut {NoStop}%
\bibitem [{\citenamefont {Holovko}, \citenamefont {Badiali},\ and\
  \citenamefont {di~Caprio}(2007)}]{HolovkoBadialidiCaprio2007}%
  \BibitemOpen
  \bibfield  {author} {\bibinfo {author} {\bibfnamefont {M.}~\bibnamefont
  {Holovko}}, \bibinfo {author} {\bibfnamefont {J.}~\bibnamefont {Badiali}}, \
  and\ \bibinfo {author} {\bibfnamefont {D.}~\bibnamefont {di~Caprio}},\
  }\href@noop {} {\bibfield  {journal} {\bibinfo  {journal} {J. Chem. Phys.}\
  }\textbf {\bibinfo {volume} {127}},\ \bibinfo {pages} {014106} (\bibinfo
  {year} {2007})}\BibitemShut {NoStop}%
\bibitem [{\citenamefont {Holovko}\ and\ \citenamefont
  {di~Caprio}(2008)}]{HolovkodiCaprio2008}%
  \BibitemOpen
  \bibfield  {author} {\bibinfo {author} {\bibfnamefont {M.}~\bibnamefont
  {Holovko}}\ and\ \bibinfo {author} {\bibfnamefont {D.}~\bibnamefont
  {di~Caprio}},\ }\href@noop {} {\bibfield  {journal} {\bibinfo  {journal} {J.
  Chem. Phys.}\ }\textbf {\bibinfo {volume} {128}},\ \bibinfo {pages} {174702}
  (\bibinfo {year} {2008})}\BibitemShut {NoStop}%
\bibitem [{\citenamefont {Jerome}(1991)}]{Jerome1991}%
  \BibitemOpen
  \bibfield  {author} {\bibinfo {author} {\bibfnamefont {B.}~\bibnamefont
  {Jerome}},\ }\href@noop {} {\bibfield  {journal} {\bibinfo  {journal} {Rep.
  Prog. Phys.}\ }\textbf {\bibinfo {volume} {54}},\ \bibinfo {pages} {391}
  (\bibinfo {year} {1991})}\BibitemShut {NoStop}%
\bibitem [{\citenamefont {Carnie}\ and\ \citenamefont
  {Chan}(1981)}]{CarnieChan1981}%
  \BibitemOpen
  \bibfield  {author} {\bibinfo {author} {\bibfnamefont {S.}~\bibnamefont
  {Carnie}}\ and\ \bibinfo {author} {\bibfnamefont {D.}~\bibnamefont {Chan}},\
  }\href@noop {} {\bibfield  {journal} {\bibinfo  {journal} {J. Chem. Phys.}\
  }\textbf {\bibinfo {volume} {74}},\ \bibinfo {pages} {1293} (\bibinfo {year}
  {1981})}\BibitemShut {NoStop}%
\bibitem [{Note1()}]{Note1}%
  \BibitemOpen
  \bibinfo {note} {Note that $S_{l\mu } (z,\Omega _{wn})$ for $\mu \protect
  \neq 0$ is complex, it is usual to introduce the following real quantities
  $S_{l\mu }^+ = (S_{l\mu } + S_{l\mu }^*)/2$ and $S_{l\mu }^- = (S_{l\mu } -
  S_{l\mu }^*)/(2i)$ where * denotes the complex conjugate.}\BibitemShut
  {Stop}%
\bibitem [{\citenamefont {Holovko}, \citenamefont {Kravtsiv},\ and\
  \citenamefont {di~Caprio}(2013)}]{HolovkoKravtsivdiCaprio2013}%
  \BibitemOpen
  \bibfield  {author} {\bibinfo {author} {\bibfnamefont {M.}~\bibnamefont
  {Holovko}}, \bibinfo {author} {\bibfnamefont {I.}~\bibnamefont {Kravtsiv}}, \
  and\ \bibinfo {author} {\bibfnamefont {D.}~\bibnamefont {di~Caprio}},\
  }\href@noop {} {\bibfield  {journal} {\bibinfo  {journal} {Cond. Matter.
  Phys.}\ }\textbf {\bibinfo {volume} {16}},\ \bibinfo {pages} {14002}
  (\bibinfo {year} {2013})}\BibitemShut {NoStop}%
\bibitem [{\citenamefont {Maier}\ and\ \citenamefont
  {Saupe}(1959)}]{MaierSaupe1959}%
  \BibitemOpen
  \bibfield  {author} {\bibinfo {author} {\bibfnamefont {W.}~\bibnamefont
  {Maier}}\ and\ \bibinfo {author} {\bibfnamefont {A.}~\bibnamefont {Saupe}},\
  }\href@noop {} {\bibfield  {journal} {\bibinfo  {journal} {Z. Naturforsch.}\
  }\textbf {\bibinfo {volume} {A14}},\ \bibinfo {pages} {882} (\bibinfo {year}
  {1959})}\BibitemShut {NoStop}%
\bibitem [{\citenamefont {Maier}\ and\ \citenamefont
  {Saupe}(1960)}]{MaierSaupe1960}%
  \BibitemOpen
  \bibfield  {author} {\bibinfo {author} {\bibfnamefont {W.}~\bibnamefont
  {Maier}}\ and\ \bibinfo {author} {\bibfnamefont {A.}~\bibnamefont {Saupe}},\
  }\href@noop {} {\bibfield  {journal} {\bibinfo  {journal} {Z. Naturforsch.}\
  }\textbf {\bibinfo {volume} {A15}},\ \bibinfo {pages} {287} (\bibinfo {year}
  {1960})}\BibitemShut {NoStop}%
\bibitem [{\citenamefont {Sokolovska}, \citenamefont {Sokolovskii},\ and\
  \citenamefont {Patey}(2004)}]{Sokolovska2004}%
  \BibitemOpen
  \bibfield  {author} {\bibinfo {author} {\bibfnamefont {T.}~\bibnamefont
  {Sokolovska}}, \bibinfo {author} {\bibfnamefont {R.}~\bibnamefont
  {Sokolovskii}}, \ and\ \bibinfo {author} {\bibfnamefont {G.}~\bibnamefont
  {Patey}},\ }\href@noop {} {\bibfield  {journal} {\bibinfo  {journal} {Phys.
  Rev. Lett.}\ }\textbf {\bibinfo {volume} {92}},\ \bibinfo {pages} {185508}
  (\bibinfo {year} {2004})}\BibitemShut {NoStop}%
\bibitem [{\citenamefont {Sokolovska}, \citenamefont {Sokolovskii},\ and\
  \citenamefont {Patey}(2005)}]{Sokolovska2005}%
  \BibitemOpen
  \bibfield  {author} {\bibinfo {author} {\bibfnamefont {T.}~\bibnamefont
  {Sokolovska}}, \bibinfo {author} {\bibfnamefont {R.}~\bibnamefont
  {Sokolovskii}}, \ and\ \bibinfo {author} {\bibfnamefont {G.}~\bibnamefont
  {Patey}},\ }\href@noop {} {\bibfield  {journal} {\bibinfo  {journal} {J.
  Chem. Phys.}\ }\textbf {\bibinfo {volume} {122}},\ \bibinfo {pages} {034703}
  (\bibinfo {year} {2005})}\BibitemShut {NoStop}%
\bibitem [{\citenamefont {Kravtsiv}\ \emph {et~al.}()\citenamefont {Kravtsiv},
  \citenamefont {Patsahan}, \citenamefont {Holovko},\ and\ \citenamefont
  {di~Caprio}}]{KravtsivPatsahanHolovkodiCaprio2015}%
  \BibitemOpen
  \bibfield  {author} {\bibinfo {author} {\bibfnamefont {I.}~\bibnamefont
  {Kravtsiv}}, \bibinfo {author} {\bibfnamefont {T.}~\bibnamefont {Patsahan}},
  \bibinfo {author} {\bibfnamefont {M.}~\bibnamefont {Holovko}}, \ and\
  \bibinfo {author} {\bibfnamefont {D.}~\bibnamefont {di~Caprio}},\ }\href@noop
  {} {\bibinfo  {journal} {in preparation}\ }\BibitemShut {NoStop}%
\end{thebibliography}%

\end{document}